\documentclass[aps,prb,10pt,twocolumn,showpacs,showkeys,citeautoscript]{revtex4-1}

\usepackage{graphicx}
\usepackage{amssymb}
\usepackage{amsmath}

\newcommand{\ud}{\,{\mathrm d}}
\newcommand{\uz}{z}
\newcommand{\uzc}{\overline{z}}
\newcommand{\uw}{w}
\newcommand{\uwc}{\overline{w}}
\newcommand{\uf}{f}
\newcommand{\ufc}{\overline{f}}
\newcommand{\zc}{\overline{z}}

\begin{document}

\title{Vortex mechanics in planar nano-magnets}

\author{Konstantin L. Metlov}
\email{metlov@fti.dn.ua}
\affiliation{Donetsk Institute for Physics and Technology NAS, Donetsk, Ukraine 83114}
\date{\today}
\begin{abstract}
A collective-variable approach for the study of non-linear dynamics of
magnetic textures in planar nano-magnets is proposed. The variables
are just arbitrary parameters (complex or real) in the specified
analytical function of a complex variable, describing the texture in
motion. Starting with such a function, a formal procedure is outlined,
allowing a (non-linear) system of differential equations
of motion to be obtained for the variables. The resulting equations are equivalent to
Landau-Lifshitz-Gilbert dynamics as far as the definition of collective
variables allows it. Apart from the collective-variable specification, the
procedure does not involve any additional assumptions (such as
translational invariance or steady-state motion). As an example, the
equations of weakly non-linear motion of a magnetic vortex are derived
and solved analytically. A simple formula for the dependence of the vortex
precession frequency on its amplitude is derived. The results are
verified against special cases from the literature and agree
quantitatively with experiments and simulations.
\end{abstract}
\maketitle

The idea that particles are just stable non-linear excitations of
fields is a cornerstone of modern field
theory\cite{Mie1912a,YangMills1954,Hooft2007}. The simplest such
particle, the hedgehog, was discovered theoretically by
Skyrme\cite{S58} as a solution of non-linear field equations. Many
similar configurations (topological solitons, or skyrmions) can be
supported by a variety of fields, including those with a vectorial order
parameter: magnetization, superfluid flow, or complex order parameters
in superconductors. This makes condensed matter and, especially,
magnetism a convenient setting for their study. In magnetism this
subject gained new attention following direct experimental
observations of magnetic vortices\cite{SOHSO00,WWBPMW02} and skyrmion
lattices\cite{MuhlbauerSkyrmionLattice,
  yu2010realspaceobservation}. Skyrmions are especially common
in planar nano-magnets, where they can be concisely described by
functions of complex variables\cite{M10}, parametrized by the coordinates
of their centers. Here a theoretical approach is presented for
deriving the dynamical equations of motion of skyrmion's generalized
coordinates, which are equivalent (apart from the definition of the coordinates) to
the Landau-Lifshitz dynamical equations for the magnetization vector
field\cite{LL35}. It allows the application of classical mechanics to the
complex non-linear problem of the dynamics of magnetic skyrmions, treating
them like particles.

Magnetic textures of ferromagnetic thin films usually consist of
magnetic domains\cite{Hubert_Shafer} with largely uniform
magnetization, separated by domain walls\cite{Hubert_book_walls},
where the magnetization continuously rotates between the directions of adjacent
domains. They were extensively studied in the framework of
micromagnetics and are still interesting from both fundamental and
applied points of view. The dynamics of these textures consists of
the translation of magnetic domain walls and is traditionally described by the
Thiele equation\cite{Thiele1973}, which further assumes that the
translation is steady. The possibility of such motion is a natural
assumption for magnetic textures of infinite thin films.

In laterally confined nano-magnets the textures are very
different\cite{UP93,M01_solitons2,M10}, consisting of skyrmions
(magnetic vortices and anti-vortices), some of which can be bound to
the surface\cite{M01_solitons2}. There is no translational invariance,
which makes direct application of the Thiele equation to these systems
doubtful. It still can be applied to larger
nano-magnets\cite{Huber1982, GHKDB06}, where lateral confinement is
less important, but to fully appreciate the specifics of nano-magnetism
(magnetism on sub-micron scales) a different approach is needed.

Besides the Thiele equation there are other approaches to the problem of
magnetization dynamics in laterally constrained magnets. One is based
on volume averaging of the Landau-Lifshitz-Gilbert (LLG) equation in
vector form and produces results in qualitative agreement with
micromagnetic simulations and experiments\cite{UK02,TCCGBT08}. But
it leads to an underestimate of the texture mobility\cite{TCCGBT08} and
vortex precession frequency\cite{UK02}, because the LLG equation is
non-linear (due to the constraint on the length of the magnetization vector,
masked when the equation is written in vector form) and its volume averaging (a
form of linear superposition) is, generally, not
justified. Consideration of spin-waves on a magnetic vortex
background\cite{ISMW98,ZIPC05} does reproduce the translation of the magnetic
vortex and predicts higher-energy spin-wave modes, but is limited to
linear consideration of small vortex displacements only. Including
higher order terms in the deviation of the magnetization from the (magnetic
vortex) background is not only mathematically hard, but also bound to
have difficulties reproducing the complicated non-linear motion of the
multi-vortex texture, which may completely depart from the original
static background. The non-uniform background also makes it difficult to
deal with non-local dipolar forces, which is the reason why in many of
such works (Refs~\onlinecite{ISMW98,ZIPC05} in particular) the dipolar
interaction is replaced by a local in-plane anisotropy, making the
equations partial differential (instead of integral partial
differential). This approximation is justified in the limit of
vanishing thickness of the nano-element, but quantitative agreement with
experiments and simulations in a wide range of geometries is
possible only when the magnetostatic interaction is fully accounted for.

Here an approach to magnetization dynamics in planar nano-dots is
proposed. It is a collective-variable theory, capable of dealing with
complex multi-vortex configurations (fully describing the relative motion
and deformation of the constituent vortices). Its only approximation lies
in the definition of collective variables, which are just arbitrary
parameters in a complex function of a complex variable. Given such a
function, the approach produces a system of ordinary differential
equations (ODEs) of motion (with no integral terms even in the
presence of magnetostatic interaction) for these variables. It assumes
neither translational invariance nor steady-state motion and is fully
capable of describing non-linear dynamics (if one can solve the
resulting non-linear ODEs). The external field and other potential
energy terms can be easily added without sacrificing simplicity
(the dynamical equations still remain ODEs). It also allows the inclusion of
phenomenological dissipation, akin to Gilbert's damping term in
the LLG equation. As an illustration, the equations of motion for linear and weakly
non-linear magnetic vortex dynamics in circular nano-dots are derived
and solved. As a check, the well-known analytical result for the vortex
precession frequency (originally obtained by solving the Thiele
equation) is then recovered in the limiting case of large dots.

The original motivation for this work comes from a recently published
description of low-energy (single- and multiple-vortex) magnetic
configurations in planar nano-dots of arbitrary shape in terms of
functions of a complex variable\cite{M10}. The present approach can be
thought of as a way to ``animate'' these configurations by making them
move in accordance with LLG dynamics. Despite this, one may easily
generalize it to other sets of trial functions without a substantial
sacrifice in simplicity.

The usual starting point for consideration of magnetization dynamics
is the LLG equation\cite{LL35}. It is well suited for numerical
computations, but is not good for analytical ones. This is because the
non-uniform effective field in the LLG approach, around which the magnetization vectors
precess, depends, in turn, on the whole magnetization vector field (if
the dipolar interaction is properly taken into account). This makes it
a non-linear integral partial differential equation, which is extremely
hard to solve analytically. Therefore, instead of the LLG equation, let us go
deeper and consider, as a starting point, the kinetic Lagrangian density,
which was first introduced by D\"oring\cite{Doering48}:
\begin{equation}
  \label{eq:Lagr_angular}
  \tau = - \frac{M_S}{\gamma} \left(\cos \theta - \cos \theta_0
  \right)\frac{\partial \varphi}{\partial t},
\end{equation}
where $\theta$ and $\varphi$ are the polar and azimuthal angle of the
magnetic moment $\vec{M}$ in a spherical coordinate system, $t$ is
time, $\gamma\simeq 1.76 \times 10^{11}
\mathrm{rad}/(\mathrm{s}\,\mathrm{T})$ is the gyromagnetic ratio, $M_S$ is
the saturation magnetization, and $\theta_0$ is a constant. The parametrization
of $\vec{M}$ via spherical angles conveniently satisfies the
constraint $|\vec{M}|=M_S$, leaving only two of its components
independent (and bounded). The system of two Euler-Lagrange equations
for the extremum of the corresponding action over $\theta$ and
$\varphi$ with additional time-independent potential energy terms
subtracted, is equivalent to the Landau-Lifshitz
equation\cite{Doering48,Hubert_book_walls}. The equations do not depend on
$\theta_0$, which can be used to ensure that $\tau$ is zero at the
boundary of the magnet.

The collective variable approach is then similar to the Ritz
method\cite{Ritz09} of solving boundary value problems: first, one
selects a trial function, parametrizing a wide set of possible
solutions, and then finds the values of the parameters giving the correct
answer (extremalizing a certain functional, as per the variational
principle). The Ritz method in its original formulation finds wide
applications in micromagnetics for solving static problems. Dynamics
is not much different. One may look for the extremum of action
of the full Lagrangian, including the kinetic and potential energy,
parametrized by a certain set of scalar parameters. The condition for
this extremum produces dynamical equations for the parameters,
allowing computation of their evolution in time.

While the above general recipe is applicable to an arbitrary choice of
trial functions, to make further consideration more specific, let us
focus on a particular very general family\cite{M10}. Consider a
cylinder, shown in Fig.~1, made of soft ferromagnetic material, 
with a Cartesian coordinate system, chosen in such a way that 
the axis $Z$ is perpendicular to the cylinder face ${\cal D}$, 
which is not necessarily circular.

\begin{figure}
\label{fig:illustration}
\includegraphics[scale=0.63]{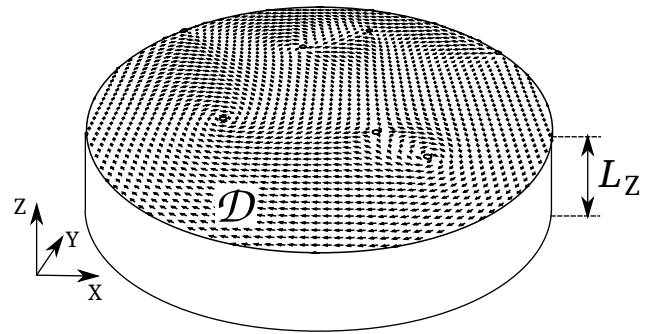}
\caption{Illustration of a ferromagnetic cylinder with coordinate
system axes (note that the present consideration is not limited to
circular cylinders; the cylinder face ${\cal D}$ can have
arbitrary shape). The arrows on top show a sample magnetic
texture $\vec{m}(\vec{r})$ from Ref.~\onlinecite{M10}, containing vortices
and anti-vortices. The positions of their centers are specified by 
collective coordinates.}
\end{figure} 
If the cylinder is sufficiently thin, the equilibrium
distribution of the magnetization vector $\vec{M}(\vec{r})$,
$\vec{r}=\{X,Y,Z\}$, inside can be assumed to be independent of the
coordinate $Z$. It can be conveniently parametrized by a complex
function $\uw(\uz,\uzc)$ of the complex variable $\uz=X+\imath Y$ (the overbar
denotes the the complex conjugation, so that $\uzc=X - \imath Y$), 
expressing the normalized magnetization $\vec{m}=\vec{M}/M_S$ as
\begin{eqnarray}
  \label{eq:mxy_w}
  m_X+\imath m_Y & = & \frac {2 \uw}{1+\uw\uwc} \\
  \label{eq:mz_w}
  m_Z & = & \pm \frac {1-\uw\uwc}{1+\uw\uwc},
\end{eqnarray}
which automatically satisfies the constraint $|\vec{m}|=1$. The sign
of $m_Z$ controls the polarization of the vortex core: $m_Z=\pm 1$ at the
vortex center.

In a flat cylinder the equilibrium static magnetization distributions
can be represented as a combination of a soliton and a meron\cite{M10}
\begin{equation}
  \label{eq:sol_SM2}
  \uw(\uz,\uzc)=\left\{
    \begin{array}{ll}
      \uf(\uz)/c_1 & |\uf(\uz)| \leq c_1 \\
      \uf(\uz)/\sqrt{\uf(\uz) \ufc(\uzc)} & c_1<|\uf(\uz)| \leq c_2\\
      \uf(\uz)/c_2 & |\uf(\uz)| > c_2,
    \end{array}
    \right. ,
\end{equation}
where $\uf(\uz)$ is an analytic [$\partial \uf(\uz)/\partial \uzc=0$]
function of the complex variable $\uz$ and $c_1$ and $c_2$ are real
constants.

In addition to the complex variable $\uz$ the function $\uw(\uz,\uzc)$
usually depends on a number of other variables like $c_1$ and $c_2$, and
others hidden inside $\uf(\uz)$. For example, the simplest 
translationally-invariant single magnetic vortex (Usov's ansatz\cite{UP93})
corresponds to
\begin{equation}
  \label{eq:usov}
  f_{\mathrm U}(z) = \frac{i (z - A)}{r_V},
\end{equation}
with $c_1=1$ (absorbed into the vortex core radius $r_V$) and
$c_2\rightarrow\infty$ in (\ref{eq:sol_SM2}), since there are no
anti-vortices. This magnetization distribution depends on the real
parameter $r_V$ and the complex parameter $A$, which are the collective
variables in this case. The parameter $A=a_X+\imath a_Y$ is the
position of the vortex center. The problem, considered below, is how
to find dynamical equations for these (and other similar) collective
variables, assuming they are functions of time
$t$, $r_V=r_V(t)$, $A=A(t)$.

The Landau-Lifshitz equation can also be written directly in complex
notation\cite{Skrotskii84}. However, as discussed earlier, our
starting point will be the kinetic part of the magnetic Lagrangian
density (\ref{eq:Lagr_angular}), expressed through the collective
variables and their time derivatives. In the complex notation its
ingredients are
 \begin{eqnarray}
   \label{eq:costheta_compl}
   \cos \theta & = & \pm \frac{1-\uw\uwc}{1+\uw\uwc} \\
   \label{eq:phi_compl}
   \exp(\imath \varphi) & = & \sqrt{\frac{\uw}{\uwc}} \\
   \label{eq:phi_partial_t_compl}
   \frac{\partial \varphi}{\partial t} & = & - \frac{\imath}{2} 
   \frac{\partial}{\partial t} \ln \frac{\uw}{\uwc},
 \end{eqnarray}
so that (\ref{eq:Lagr_angular}) can be rewritten as
\begin{eqnarray}
  \tau & = & 
  % \gamma \frac{\imath}{2} \frac{1- \uw \uwc}{1 + \uw \uwc} 
  % \,\frac{\ud}{\ud t} \ln \frac{\uw} {\uwc} = \\
  % & & \gamma \frac{\imath}{2} \frac{1- \uw \uwc}{1 + \uw \uwc} \,
  % \left(
  %  \frac{\dot{\uw}} {\uw} -
  %  \frac{\dot{\uwc}} {\uwc}
  %\right) =
  \mp \frac{M_S}{\gamma} {\mathrm{Im}} 
  \frac{1- \uw \uwc}{1 + \uw \uwc} \frac{\dot{\uw}} {\uw},
  \label{eq:Lagr_complex}
\end{eqnarray}
where the dot over a variable denotes the time derivative and
$\uw(\uz,\uzc,t)$ from (\ref{eq:sol_SM2}) depends on time $t$ via the
collective variables in the trial function $\uf(\uz)$. The meron part
of (\ref{eq:sol_SM2}), as per our selection of $\theta_0=\pi/2$, gives
no contribution to the kinetic Lagrangian, because $|\uw(\uz,\uzc)|=1$
inside it, and therefore $\tau \sim m_Z \sim \cos\theta= 0$.

To derive the dynamical equations the total Lagrangian is needed, which is the
Lagrangian density integrated over the particle volume
\begin{equation}
  \label{eq:Lagr_tot_def}
  T = \int_{{\cal D}\times L_Z} \tau \ud^3\vec{r}  = 
          L_Z \int_{{\cal D}_S} \tau \ud^2\vec{r},
\end{equation}
where $L_Z$ is the cylinder's thickness and ${\cal D}_S$ is the part of
the cylinder's face, occupied by soliton (\ref{eq:sol_SM2}), for which
$|\uf(\uz)|<c_1$ or $|\uf(\uz)|>c_2$. It can be simplified for
arbitrary $\uf$ in (\ref{eq:sol_SM2}) by noting that
\begin{equation}
  \label{eq:ident}
  \frac{\partial}{\partial t} 
  \ln \frac{4 \uw(t)\uwc(t')}{(1+\uw(t)\uwc(t'))^2}=
  \frac{1- \uw(t) \uwc(t')}{1 + \uw(t) \uwc(t')}
  \frac{1} {\uw(t)} \frac{\partial \uw(t)}{\partial t}
\end{equation}
where the variable $t'$ is considered independent and does not take
part in the differentiation. Interchanging the ${\mathrm{Im}}$ operation
and the time derivative with the area integral (which is possible because
the area element is real and integration is a linear operation) we
arrive at the following expression for the total kinetic Lagrangian
\begin{equation}
  \label{eq:Lagr_tot_def1}
  T = \mp \frac{M_S L_Z}{\gamma}
  \left.
    {\mathrm{Im}} \frac {\partial}{\partial t} 
    \!\!\int_{{\cal D}_S'} \!\!\!\!\!\ln 
    \frac{4\, \uw(z, t) \uwc(\zc, t')}
    {(1+\uw(z,t) \uwc(\zc, t'))^2} 
    \ud^2 z
  \right|_{t'\rightarrow t},
\end{equation}
where for the purpose of calculating the area integral and
differentiating it is assumed that all the collective variables inside
$\uwc$ and the definition of the integration region ${\cal D}_S'$ (note the
prime) depend on the new independent time variable $t'$, whereas
inside $\uw$ they still depend on $t$ (only after differentiation 
is $t'$ replaced by $t$). It was also noted that inside the soliton
${\cal D}_S$ the function $\uw$ is analytic and does not depend on
$\uzc$, while its conjugate $\uwc$ does not depend on $\uz$. This
formula can be checked directly. It can be further simplified if there
are no boundary-bound vortices and anti-vortices. In this case it is
possible to integrate (\ref{eq:Lagr_tot_def1}) by parts, making use of
Greene's formula
\begin{equation}
  \label{eq:Greene}
  \frac{1}{2 \imath} \oint_{\partial {\cal D}} u(\uz,\uzc) \ud \uz = 
  \iint_{\cal D} \frac{\partial u(\uz,\uzc)}{\partial \uzc} \ud^2 z,
\end{equation}
for any reasonably good function $u$, which yields
\begin{equation}
  \label{eq:Lagr_tot_def2}
  T = \pm \frac{M_S L_Z}{\gamma}
  \left[
    {\mathrm{Im}} \frac {\partial}{\partial t} 
    \int_{{\cal D}_S'} 
    \uzc 
    \frac{1-\uw \uwc}
    {1+\uw \uwc}
    \frac{1}{\uwc}
    \frac{\partial \uwc}{\partial \uzc}
    \ud^2 z
  \right]_{t'\rightarrow t},
\end{equation}
where the function arguments are omitted, but it is still assumed that all
the collective variables inside $\uw$ depend on $t$ and all of them
inside $\uwc$ and the definition of the region ${\cal D}_S'$ depend on
$t'$.

The expression (\ref{eq:Lagr_tot_def1}) (and (\ref{eq:Lagr_tot_def2})
for the case of solitons fully contained inside the particle) is the
main result of this paper. To derive the equations of motion for
the collective variables $x_i(t)$, entering the trial function
$\uf(\{x_i\},z)$, it is now sufficient to write down the full Lagrangian:
\begin{equation}
  \label{eq:Lagr_full}
  L(\{\dot{x}_i\},\{x_i\}) = T(\{\dot{x}_i\},\{x_i\}) - U(\{x_i\}),
\end{equation}
where $U$ is the potential energy (including exchange, magnetostatic,
and, possibly, other energy terms); and use it to derive the system of
Euler-Lagrange equations
\begin{equation}
  \label{eq:EulerLagrange}
  \frac{\ud}{\ud t} \frac{\partial L}{\partial \dot{x_i}}
  -
  \frac{\partial L}{\partial x_i}
   = 0,
\end{equation}
which extremize the corresponding action. This allows
problems of (multi-) vortex dynamics to be treated as problems of classical
mechanics. It is also worth noting that apart from restrictions,
implied by a particular choice of $\uf$, in defining the collective variables
$x_i$, the above consideration involves no approximations and
corresponds to the solution of the LLG equation exactly. This is
especially easy to see by considering a discrete magnet with each spin
parametrized by the spherical angles $\theta_{ijk}$ and $\varphi_{ijk}$ and
introducing discrete analogs of the static interactions (exchange,
dipolar, etc) between the spins with finite differences instead of
spatial derivatives. When all these spherical angles for each spin are
chosen as independent variables, the equations of motion
(\ref{eq:EulerLagrange}) in the continuum limit (when the number of spins goes
to infinity, while the magnet volume is constant) coincide with the
Landau-Lifshitz equation for $\theta(\vec{r},t)$ and
$\varphi(\vec{r},t)$.

Let us now proceed with examples. 

First, to illustrate that, as in classical mechanics, this
theory may suffer from poor selection of trial functions, consider
the dynamics of uniformly displaced magnetic vortex (\ref{eq:usov}) with
$r_V=\mathrm{const}$ and $A=A(t)$. One may readily check that all three
expressions for the kinetic Lagrangian (\ref{eq:Lagr_tot_def}),
(\ref{eq:Lagr_tot_def1}), and (\ref{eq:Lagr_tot_def2}) yield 0 in this
case. This means that the Euler-Lagrange equations
(\ref{eq:EulerLagrange}) reduce to conditions of static equilibrium,
or (in the case of the potential energy, consisting of the exchange and
magnetostatic terms) that such an undeformed vortex stays in the center
of the cylinder. This is similar to the conclusion from micromagnetics
that moving domain walls always have a different profile
from that of stationary ones\cite{Schlomann73}. In some
sense, one may say that such a modification of a magnetic texture (of a domain
wall or a magnetic vortex) is the way in which it ``remembers'' that it is
moving.

Let us now turn to a more complex trial function, describing
vortex displacement without the formation of magnetic charges on the
cylinder's side\cite{M01_solitons2}, which is a particular case of
a more general class of trial functions\cite{M10}
\begin{equation}
  \label{eq:noside}
  f(z)=\imath \frac{z - (A + \overline{A} z^2)}{r_V},
\end{equation}
where again $r_V=\mathrm{const}$ and $A=A(t)=a_X(t)+\imath a_Y(t)$, $|A|<1/2$,
is a complex pair of collective coordinates. The vortex center, where
$f(z_C)=0$, is not exactly at $z_C=A$, as in the case of
a uniformly displaced vortex, but rather at $z_C=(1-\sqrt{1-4 A
  \overline{A}})/(2 \overline{A})$. While it is possible to change
variables and write the equations of motion for $z_C(t)$ directly, let
us illustrate one of the powers of the present approach, which is a
great freedom in selecting the parametrization, and write them for
$A(t)$. Also note that here the parameter $a$ in the original
expression of Ref.~\onlinecite{M01_solitons2} is substituted by
$-\imath A$, which makes the vortex center displacements coincide in
phase with the complex parameter $A$. That is, real $A$ correspond now
to real $z_C$. This again is a matter of convenience and does not
change anything, since the parametrization can be arbitrary. The total
Lagrangian up to the second order in $|A|$ from
(\ref{eq:Lagr_tot_def2}) and (\ref{eq:Lagr_full}) with full account
for the vortex core shape deformation is
\begin{equation}
  {\cal L} \!=\!
  \pm \kappa_2\left(a_X(t)\dot{a}_Y(t)\!-\!a_Y(t)\dot{a}_X(t)\right)
  - k_2\left(a_X^2(t)\!+\!a_Y^2(t)\right),
  \label{eq:Lnoside}
\end{equation}
where ${\cal L}=L/(\mu_0 M_S^2 \pi L_Z R^2)$ and $k_2$ are dimensionless
and $\kappa_2=[1+r_V^4(4 \ln\!2 - 3)]/(\gamma \mu_0 M_S)$ has units
of seconds; $k_2$ is the second order expansion coefficient of the
potential energy (consisting of exchange and dipolar terms). A constant
zero-order potential energy term, equal to the equilibrium
energy of the centered vortex, was omitted because it has no influence on
dynamics of $a(t)$. The equations of motion (\ref{eq:EulerLagrange}) are
\begin{eqnarray}
\kappa_2 \dot{a}_X(t) \pm k_2 a_Y(t) & = & 0 , \nonumber \\
\kappa_2 \dot{a}_Y(t) \mp k_2 a_X(t) & = & 0 , \label{eq:nocharges2eq}
\end{eqnarray}
which, for initial conditions $a_X(0)=a_0$, $a_Y(y)=0$, have the
following solution:
\begin{eqnarray}
a_X(t) & = & a_0 \cos(\omega t), \nonumber \\
a_Y(t) & = & \pm a_0 \sin(\omega t), \label{eq:nocharges2sol}
\end{eqnarray}
corresponding to the circular motion of the vortex around the dot
center with frequency $\omega=\omega_0=k_2/\kappa_2$. It is important
to note that the direction of vortex motion is not arbitrary. It
depends on its core polarization, but not its chirality, since $T$ and $U$
are independent of the sign of $w$ or $f$. The vortices with $m_Z=-1$
at the center rotate clockwise, and the vortices with $m_Z=1$
counterclockwise. This is in full agreement with the simulations and
experiments of Ref.~\onlinecite{Choe2004}, but in disagreement with
its conclusions, since vortex chirality (included in ``handedness'')
plays no role in determining the direction of vortex rotation. This
also allows us to guess that the vortex core polarization in the
simulation of Ref.~\onlinecite{UK02} was positive, which is natural to
assume, but was not specified by the authors. A similar polarization
of the core can be guessed from Fig.~2 in Ref.~\onlinecite{GHKDB06},
but with significant uncertainty, since it is masked by low resolution
of the measurement in the $Y$ direction, as discussed therein.

To make a more rigorous quantitative confirmation of the present
theory, let us compute the rotation frequency of the magnetic vortex in the
limit of large flat circular dots with $L_Z \ll R$ and $R \gg L_E$,
where $L_E=\sqrt{C/\mu_0 M_S^2}$ is the exchange length, $C$ is the
exchange stiffness, and $R$ is the dot's radius. In this case $r_V \ll 1$,
and $\kappa_2\simeq 1/(\gamma \mu_0 M_S)$. The second order expansion
of the energy of the vortex (\ref{eq:noside}) with the vortex core
neglected was published in Ref.~\onlinecite{MG02_JEMS}. If the
exchange contribution of order $L_E/R \ll 1$ is also neglected in that
expansion, the coefficient $k_2$ in large dots is fully determined by
the energy of the volume magnetic charges \cite{MG02_JEMS}. Converting to
SI units, for the precession frequency we get
\begin{eqnarray}
  \nu & = & \frac{k_2}{2\pi\kappa_2}=\frac{\gamma \mu_0 M_S}{\pi} \!\!\!
  \int\limits_0^\infty
  \frac{f_\mathrm{MS}(k g)}{k}
  \left[
    \int\limits_0^1 \! r J_1(k r) \ud r
  \right]^2 \!\!\!\!\ud k, \\
  \nu & \simeq & \gamma \mu_0 M_S g \frac{2(2G-1)}{6\pi^2}
  \label{eq:freqasympt}
\end{eqnarray}
where $f_\mathrm{MS}(x)=1-(1-e^{- x})/x$, $g=L_Z/R\ll 1$ and
$G\simeq0.915\,966$ is Catalan's constant. This expression [apart from
measurement units and the value of the numerical constant in
(\ref{eq:freqasympt}), which is exact here] coincides with the
expression for the vortex frequency, obtained in Ref.~\onlinecite{GHKDB06}
on the basis of the Thiele equation and quantitatively confirmed there by
experiments on large dots. It is worth noting that in
Ref.~\onlinecite{GHKDB06} different terms in the equation of motion
correspond to different models: the dynamical term with time derivatives
comes from the Thiele equation for uniform steady translation of magnetic
texture, while the potential energy term assumes the non-uniform mode of
vortex displacement from Ref.~\onlinecite{MG02_JEMS}. This is,
strictly speaking, not consistent and works only because in large dots
$r_V \ll 1$ and the dynamical term $\kappa_2 \rightarrow 1$ becomes
insensitive to the vortex core shape deformation. The derivation of the
vortex precession frequency above is fully consistent and uses the
same trial function for both the kinetic and potential energy terms.

Real magnets inevitably dissipate the energy of moving spins in the form
of heat. But the Lagrangian formalism in its pure form does not include
dissipation. It is added externally via the Rayleigh dissipation
function $D$
\begin{equation}
  D = \frac{1}{2} \sum_i \sum_j D_{ij}\dot{x_i} \dot{x_j},
\end{equation}
which is then included into the right hand side of the Euler-Lagrange
equations (\ref{eq:EulerLagrange}) as an additional term $-\partial
D/\partial \dot{x_i}$. The matrix $D_{ij}$ consists of
phenomenological dissipation coefficients. Judging from the abstract
of the unpublished report by Gilbert\cite{Gilbert55} it is possible to
speculate that the Lagrangian formalism was also his starting point and
his dissipative term (whose full microscopic justification is still an
open problem\cite{Kambersky2007}) has similar origins. Thus, $D_{ij}$
must be related to Gilbert's phenomenological dissipation
constant. This relation is, probably, best established by considering
the energy balance in the system, but let us leave it for now as an
open problem and treat $D_{ij}$ as independent phenomenological
parameters. A choice of $D=d [\dot{a}_X(t)^2+\dot{a}_Y(t)^2]$ changes the
solution (\ref{eq:nocharges2sol}) into
\begin{eqnarray}
a_X(t) & = & a_0 e^{-d k_2 t/(d^2+\kappa_2^2)} 
                    \cos\left[k_2 t \kappa_2/(d^2+\kappa_2^2)\right], \nonumber \\
a_Y(t) & = & \pm a_0 e^{-d k_2 t/(d^2+\kappa_2^2)} 
                    \sin\left[k_2 t \kappa_2/(d^2+\kappa_2^2)\right]. \label{eq:nocharges2soldiss}
\end{eqnarray}
As one can see, as in the case of a linear oscillator, the vortex
precession frequency starts to depend (slightly) on a (small) damping
coefficient.

Finally, let us consider weakly non-linear vortex dynamics by taking
into account the kinetic and potential energy terms, corresponding to
the fourth order in $|A|$. Continuing the expansion of the kinetic
Lagrangian (\ref{eq:Lagr_tot_def2}) with the trial function
(\ref{eq:noside}) leads to the following expression:
\begin{eqnarray}
  {\cal L} & = &
  \pm \{ \kappa_2 + \kappa_4 [a_X^2(t)\!+\!a_Y^2(t)]\}\left[a_X(t)\dot{a}_Y(t)\!-\!a_Y(t)\dot{a}_X(t)\right] \nonumber \\
  & & - k_2\left[a_X^2(t)\!+\!a_Y^2(t)\right]   
  - k_4\left[a_X^2(t)\!+\!a_Y^2(t)\right]^2,
  \label{eq:Lnoside4}  
\end{eqnarray}
where $\kappa_4=2-r_V^2\{23+r_V^2[(6061 - 6397 r_V^2)/8 -
1152(1-r_V^2)\ln 2]\}/(\gamma \mu_0 M_S)$ like $\kappa_2$ has units of
seconds and $k_4$ is the next potential energy expansion
coefficient. The corresponding equations of motion become non-linear,
but they are solved exactly by (\ref{eq:nocharges2sol}) with
\begin{equation}
  \omega =  \frac{k_2 + 2 a_0^2 k_4}{\kappa_2 + 2 a_0^2 \kappa_4}\simeq
  \omega_0 + 
  2 \frac{k_4 \kappa_2 - k_2 \kappa_4}{\kappa_2^2} a_0^2 + O(a_0^4). \label{eq:freq4}
\end{equation}
As in other non-linear oscillators, the vortex rotation
frequency becomes dependent on the rotation amplitude.  Derivation of
the expressions for $k_2$ and $k_4$ in the general case with full account
for vortex core deformation is rather cumbersome and, together with
the analysis of their dependence on the dot dimensions, will be the
subject of another forthcoming presentation. Nevertheless, preliminary
versions of these expressions are attached in the form of a {\tt MATHEMATICA}
file as a Supplemental Material\cite{suppfrequency}. They can be used to compute vortex
precession frequencies for various dot geometries, not covered here.

Limitations of the presented approach follow from its
strengths. The results and the procedure are simple, but are just as good as
the selected trial function. This is similar to the applications of
the Ritz method to static problems of micromagnetics.  Comparing the
results obtained with different trial functions for a particular
problem allows the one giving the most realistic
description to be chosen. The quantitative basis for such a comparison can be the
total action, corresponding to the evolution of the system between two
known states. There are complications, however, due to the fact that
the Lagrangian formalism prescribes that the action is stationary, but not
necessarily minimal. Thus, development of a firm basis for comparison
of different trial functions in the Lagrangian formalism might be an
interesting possibility for future research with potential benefits
across different branches of physics. In any case, more trial
functions are considered, closer are the best ones to the exact
analytical solution. Luckily, the family of trial functions from
Ref.~\onlinecite{M10} is huge and can be further
generalized\cite{MG04, M06}, which facilitates such a competition.
Vortex/anti-vortex pair nucleation is another problem, which requires
a separate treatment. New vortices (changes in the topological charge)
always come into the element through its boundary, a process
well described by the considered family of trial functions in both
the single-\cite{M01_solitons2, MG02_JEMS} and multiple-\cite{M10} vortex
cases. Vortex-antivortex annihilation is simple and corresponds to
cancellation of monomials in the numerator and denominator of a rational
complex trial function\cite{M10}. Vortex-antivortex pair nucleation
(without change in the total topological charge), however, introduces
branching in the trial functions, where at some point in time and space
additional monomials in the numerator and denominator of the rational
function appear and spread out. While further evolution of the nucleated
vortex-antivortex pair can be described by the presented approach
directly, their initial nucleation requires the above-mentioned
rigorous comparison between trial functions to detect when a trial
function with more vortices and anti-vortices should replace the
original one. Such branching points will have to be introduced into
the dynamical process externally by showing that the total action of
the process with nucleation and further evolution of the nucleated pair
somewhere along the trajectory is lower than that with
continued evolution of the original number of vortices. Consideration
of branching might require an introduction of graph techniques similar
to Feynman's diagrams\cite{Feynman1949}.

Despite its limitations, the presented Lagrangian approach to linear
and non-linear magnetic vortex dynamics can be directly applied to
many interesting and useful problems of magnetism, such as magnetic
vortex resonance in particles of various shapes (and influence of the
particle shape on its frequency); the dynamics of vortex nucleation, when
the ``C''-shaped magnetization state transforms into a vortex dynamically; the dynamics of
charged finite domain walls in nano-strips, which are well described
by complex trial functions\cite{M10}; externally driven non-linear
resonance; and chaos in unsaturated nano-magnets.

In conclusion, several equivalent alternative expressions for the kinetic
Lagrangian (\ref{eq:Lagr_complex}), (\ref{eq:Lagr_tot_def}),
(\ref{eq:Lagr_tot_def1}), and (\ref{eq:Lagr_tot_def2}) of an arbitrary
trial function $f$, defining collective variables in (possibly
multi-vortex) magnetic textures\cite{M10} in flat nano-elements are
derived. They allow  non-linear equations of motion to be obtained for these
variables similar to the ones in classical Lagrangian
mechanics. Apart from the collective variable definition, this theory is
exact and involves no additional approximations beyond those of
the Landau-Lifshitz-Gilbert equation. It is validated here by considering
magnetic vortex precession in a cylindrical nano-dot. In the limit of
large flat dots its frequency coincides with experimental data and
known theoretical estimations, based on the Thiele
equation\cite{GHKDB06}. The question of the direction of the vortex
rotation is elucidated. It is found to depend on vortex core
polarization only and not on its chirality. Also, analytical
solutions for vortex rotation in a dissipative magnet are derived
(\ref{eq:nocharges2soldiss}); its frequency is decreased by
damping. Finally, weakly non-linear rotation of the vortex is considered,
allowing the relation (\ref{eq:freq4}) between its
frequency and amplitude to be established via potential energy expansion
coefficients. The expressions for the kinetic Lagrangian in
(\ref{eq:Lnoside}) and (\ref{eq:Lnoside4}) for the trial function
(\ref{eq:noside}) can be reused in other calculations, including
evaluation of the time-dependent external field, spin torque, and other potential energy
terms. One may expect them to be as simple as the examples above.

I would like to thank Vladimir N. Krivoruchko for reading the manuscript
and many valuable suggestions.

%\bibliographystyle{apsrev4-1}
%\bibliography{klm_base}

\begin{thebibliography}{10}%
\makeatletter
\providecommand \@ifxundefined [1]{%
 \ifx #1\undefined \expandafter \@firstoftwo
 \else \expandafter \@secondoftwo
\fi
}%
\providecommand \@ifnum [1]{%
 \ifnum #1\expandafter \@firstoftwo
 \else \expandafter \@secondoftwo
\fi
}%
\providecommand \enquote [1]{``#1''}%
\providecommand \bibnamefont  [1]{#1}%
\providecommand \bibfnamefont [1]{#1}%
\providecommand \citenamefont [1]{#1}%
\providecommand\href[0]{\@sanitize\@href}%
\providecommand\@href[1]{\endgroup\@@startlink{#1}\endgroup\@@href}%
\providecommand\@@href[1]{#1\@@endlink}%
\providecommand \@sanitize [0]{\begingroup\catcode`\&12\catcode`\#12\relax}%
\@ifxundefined \pdfoutput {\@firstoftwo}{%
 \@ifnum{\z@=\pdfoutput}{\@firstoftwo}{\@secondoftwo}%
}{%
 \providecommand\@@startlink[1]{\leavevmode}%
 \providecommand\@@endlink[0]{}%
}{%
 \providecommand\@@startlink[1]{%
  \leavevmode
  \pdfstartlink
   attr{/Border[0 0 1 ]/H/I/C[0 1 1]}%
   user{/Subtype/Link/A<</Type/Action/S/URI/URI(#1)>>}%
  \relax
 }%
 \providecommand\@@endlink[0]{\pdfendlink}%
}%
\providecommand \url  [0]{\begingroup\@sanitize \@url }%
\providecommand \@url [1]{\endgroup\@href {#1}{\urlprefix}}%
\providecommand \urlprefix [0]{URL }%
\providecommand \Eprint[0]{\href }%
\@ifxundefined \urlstyle {%
  \providecommand \doi [1]{doi:\discretionary{}{}{}#1}%
}{%
  \providecommand \doi [0]{doi:\discretionary{}{}{}\begingroup
  \urlstyle{rm}\Url }%
}%
\providecommand \doibase [0]{http://dx.doi.org/}%
\providecommand \Doi[1]{\href{\doibase#1}}%
\providecommand \bibAnnote [3]{%
  \BibitemShut{#1}%
  \begin{quotation}\noindent
    \textsc{Key:}\ #2\\\textsc{Annotation:}\ #3%
  \end{quotation}%
}%
\providecommand \bibAnnoteFile [2]{%
  \IfFileExists{#2}{\bibAnnote {#1} {#2} {\input{#2}}}{}%
}%
\providecommand \typeout [0]{\immediate \write \m@ne }%
\providecommand \selectlanguage [0]{\@gobble}%
\providecommand \bibinfo [0]{\@secondoftwo}%
\providecommand \bibfield [0]{\@secondoftwo}%
\providecommand \translation [1]{[#1]}%
\providecommand \BibitemOpen[0]{}%
\providecommand \bibitemStop [0]{}%
\providecommand \bibitemNoStop [0]{.\EOS\space}%
\providecommand \EOS [0]{\spacefactor3000\relax}%
\providecommand \BibitemShut [1]{\csname bibitem#1\endcsname}%
%</preamble>
\bibitem{Mie1912a}%
  \BibitemOpen
  \bibfield{author}{%
  \bibinfo {author} {\bibfnamefont{G.}~\bibnamefont{Mie}},\ }%
  \bibfield{journal}{%
  \Doi{10.1002/andp.19123420306}{\bibinfo {journal} {Annalen der Physik}}\ }%
  \textbf{\bibinfo {volume} {342}},\ \bibinfo {pages} {511} (\bibinfo {year}
  {1912}),\ ISSN \bibinfo {issn} {1521-3889}%
  \bibAnnoteFile{NoStop}{Mie1912a}%
\bibitem{YangMills1954}%
  \BibitemOpen
  \bibfield{author}{%
  \bibinfo {author} {\bibfnamefont{C.~N.}\ \bibnamefont{Yang}}\ and\ \bibinfo
  {author} {\bibfnamefont{R.~L.}\ \bibnamefont{Mills}},\ }%
  \bibfield{journal}{%
  \Doi{10.1103/PhysRev.96.191}{\bibinfo {journal} {Phys. Rev.}}\ }%
  \textbf{\bibinfo {volume} {96}},\ \bibinfo {pages} {191} (\bibinfo {month}
  {Oct}\ \bibinfo {year} {1954})%
  \bibAnnoteFile{NoStop}{YangMills1954}%
\bibitem{Hooft2007}%
  \BibitemOpen
  \bibfield{author}{%
  \bibinfo {author} {\bibfnamefont{G.}~\bibnamefont{{'t Hooft}}},\ }%
  in\ \emph{\bibinfo {booktitle} {{Philosophy of Physics, Part A.}}},\ \bibinfo
  {editor} {edited by\ \bibinfo {editor}
  {\bibfnamefont{J.}~\bibnamefont{Butterfield}}\ and\ \bibinfo {editor}
  {\bibfnamefont{J.}~\bibnamefont{Earman}}}\ (\bibinfo {publisher} {Elsevier},\
  \bibinfo {year} {2007})\ pp.\ \bibinfo {pages} {661--730}%
  \bibAnnoteFile{NoStop}{Hooft2007}%
\bibitem{S58}%
  \BibitemOpen
  \bibfield{author}{%
  \bibinfo {author} {\bibfnamefont{T.~H.~R.}\ \bibnamefont{Skyrme}},\ }%
  \bibfield{journal}{%
  \bibinfo {journal} {Proc. Roy. Soc. A}\ }%
  \textbf{\bibinfo {volume} {247}},\ \bibinfo {pages} {260} (\bibinfo {year}
  {1958})%
  \bibAnnoteFile{NoStop}{S58}%
\bibitem{SOHSO00}%
  \BibitemOpen
  \bibfield{author}{%
  \bibinfo {author} {\bibfnamefont{T.}~\bibnamefont{Shinjo}}, \bibinfo {author}
  {\bibfnamefont{T.}~\bibnamefont{Okuno}}, \bibinfo {author}
  {\bibfnamefont{R.}~\bibnamefont{Hassdorf}}, \bibinfo {author}
  {\bibfnamefont{K.}~\bibnamefont{Shigeto}},\ and\ \bibinfo {author}
  {\bibfnamefont{T.}~\bibnamefont{Ono}},\ }%
  \bibfield{journal}{%
  \bibinfo {journal} {Science}\ }%
  \textbf{\bibinfo {volume} {289}},\ \bibinfo {pages} {930} (\bibinfo {year}
  {2000})%
  \bibAnnoteFile{NoStop}{SOHSO00}%
\bibitem{WWBPMW02}%
  \BibitemOpen
  \bibfield{author}{%
  \bibinfo {author} {\bibfnamefont{A.}~\bibnamefont{Wachowiak}}, \bibinfo
  {author} {\bibfnamefont{J.}~\bibnamefont{Wiebe}}, \bibinfo {author}
  {\bibfnamefont{M.}~\bibnamefont{Bode}}, \bibinfo {author}
  {\bibfnamefont{O.}~\bibnamefont{Pietzsch}}, \bibinfo {author}
  {\bibfnamefont{M.}~\bibnamefont{Morgenstern}},\ and\ \bibinfo {author}
  {\bibfnamefont{R.}~\bibnamefont{Wiesendanger}},\ }%
  \bibfield{journal}{%
  \Doi{10.1126/science.1075302}{\bibinfo {journal} {{Science}}}\ }%
  \textbf{\bibinfo {volume} {298}},\ \bibinfo {pages} {577} (\bibinfo {year}
  {2002})%
  \bibAnnoteFile{NoStop}{WWBPMW02}%
\bibitem{MuhlbauerSkyrmionLattice}%
  \BibitemOpen
  \bibfield{author}{%
  \bibinfo {author} {\bibfnamefont{S.}~\bibnamefont{M{\"u}hlbauer}}, \bibinfo
  {author} {\bibfnamefont{B.}~\bibnamefont{Binz}}, \bibinfo {author}
  {\bibfnamefont{F.}~\bibnamefont{Jonietz}}, \bibinfo {author}
  {\bibfnamefont{C.}~\bibnamefont{Pfleiderer}}, \bibinfo {author}
  {\bibfnamefont{A.}~\bibnamefont{Rosch}}, \bibinfo {author}
  {\bibfnamefont{A.}~\bibnamefont{Neubauer}}, \bibinfo {author}
  {\bibfnamefont{R.}~\bibnamefont{Georgii}},\ and\ \bibinfo {author}
  {\bibfnamefont{P.}~\bibnamefont{B{\"o}ni}},\ }%
  \bibfield{journal}{%
  \Doi{10.1126/science.1166767}{\bibinfo {journal} {Science}}\ }%
  \textbf{\bibinfo {volume} {323}},\ \bibinfo {pages} {915} (\bibinfo {year}
  {2009})%
  \bibAnnoteFile{NoStop}{MuhlbauerSkyrmionLattice}%
\bibitem{yu2010realspaceobservation}%
  \BibitemOpen
  \bibfield{author}{%
  \bibinfo {author} {\bibfnamefont{X.}~\bibnamefont{Yu}}, \bibinfo {author}
  {\bibfnamefont{Y.}~\bibnamefont{Onose}}, \bibinfo {author}
  {\bibfnamefont{N.}~\bibnamefont{Kanazawa}}, \bibinfo {author}
  {\bibfnamefont{J.}~\bibnamefont{Park}}, \bibinfo {author}
  {\bibfnamefont{J.}~\bibnamefont{Han}}, \bibinfo {author}
  {\bibfnamefont{Y.}~\bibnamefont{Matsui}}, \bibinfo {author}
  {\bibfnamefont{N.}~\bibnamefont{Nagaosa}},\ and\ \bibinfo {author}
  {\bibfnamefont{Y.}~\bibnamefont{Tokura}},\ }%
  \bibfield{journal}{%
  \Doi{10.1038/nature09124}{\bibinfo {journal} {Nature}}\ }%
  \textbf{\bibinfo {volume} {465}},\ \bibinfo {pages} {901} (\bibinfo {year}
  {2010})%
  \bibAnnoteFile{NoStop}{yu2010realspaceobservation}%
\bibitem{M10}%
  \BibitemOpen
  \bibfield{author}{%
  \bibinfo {author} {\bibfnamefont{K.~L.}\ \bibnamefont{Metlov}},\ }%
  \bibfield{journal}{%
  \bibinfo {journal} {Phys. Rev. Lett.}\ }%
  \textbf{\bibinfo {volume} {105}},\ \bibinfo {pages} {107201} (\bibinfo {year}
  {2010})%
  \bibAnnoteFile{NoStop}{M10}%
\bibitem{LL35}%
  \BibitemOpen
  \bibfield{author}{%
  \bibinfo {author} {\bibfnamefont{L.~D.}\ \bibnamefont{Landau}}\ and\ \bibinfo
  {author} {\bibfnamefont{E.~M.}\ \bibnamefont{Lifshitz}},\ }%
  \bibfield{journal}{%
  \bibinfo {journal} {Physik. Z. Sowjetunion}\ }%
  \textbf{\bibinfo {volume} {8}},\ \bibinfo {pages} {153} (\bibinfo {year}
  {1935})%
  \bibAnnoteFile{NoStop}{LL35}%
\bibitem{Hubert_Shafer}%
  \BibitemOpen
  \bibfield{author}{%
  \bibinfo {author} {\bibfnamefont{A.}~\bibnamefont{Hubert}}\ and\ \bibinfo
  {author} {\bibfnamefont{R.}~\bibnamefont{Sch{\"a}fer}},\ }%
  \emph{\bibinfo {title} {{Magnetic Domains. The Analysis of Magnetic
  Microstructures}}}\ (\bibinfo {publisher} {Springer},\ \bibinfo {address}
  {Berlin},\ \bibinfo {year} {1998})%
  \bibAnnoteFile{NoStop}{Hubert_Shafer}%
\bibitem{Hubert_book_walls}%
  \BibitemOpen
  \bibfield{author}{%
  \bibinfo {author} {\bibfnamefont{A.}~\bibnamefont{Hubert}},\ }%
  \emph{\bibinfo {title} {{Theorie der Dom{\"a}nenw{\"a}nde in geordneten
  Medien}}}\ (\bibinfo {publisher} {Springer},\ \bibinfo {address}
  {Berlin-Heidelberg-New York},\ \bibinfo {year} {1974})%
  \bibAnnoteFile{NoStop}{Hubert_book_walls}%
\bibitem{Thiele1973}%
  \BibitemOpen
  \bibfield{author}{%
  \bibinfo {author} {\bibfnamefont{A.~A.}\ \bibnamefont{Thiele}},\ }%
  \bibfield{journal}{%
  \Doi{10.1103/PhysRevLett.30.230}{\bibinfo {journal} {Phys. Rev. Lett.}}\ }%
  \textbf{\bibinfo {volume} {30}},\ \bibinfo {pages} {230} (\bibinfo {year}
  {1973})%
  \bibAnnoteFile{NoStop}{Thiele1973}%
\bibitem{UP93}%
  \BibitemOpen
  \bibfield{author}{%
  \bibinfo {author} {\bibfnamefont{N.~A.}\ \bibnamefont{Usov}}\ and\ \bibinfo
  {author} {\bibfnamefont{S.~E.}\ \bibnamefont{Peschany}},\ }%
  \bibfield{journal}{%
  \bibinfo {journal} {{J. Magn. Magn. Mater.}}\ }%
  \textbf{\bibinfo {volume} {118}},\ \bibinfo {pages} {L290} (\bibinfo {year}
  {1993})%
  \bibAnnoteFile{NoStop}{UP93}%
\bibitem{M01_solitons2}%
  \BibitemOpen
  \bibfield{author}{%
  \bibinfo {author} {\bibfnamefont{K.~L.}\ \bibnamefont{Metlov}},\ }%
  \enquote{\bibinfo {title} {{Two-dimensional topological solitons in soft
  ferromagnetic cylinders}},}\  (\bibinfo {year} {2001}),\ \bibinfo {note}
  {{\tt arXiv:cond-mat/0102311}}%
  \bibAnnoteFile{NoStop}{M01_solitons2}%
\bibitem{Huber1982}%
  \BibitemOpen
  \bibfield{author}{%
  \bibinfo {author} {\bibfnamefont{D.~L.}\ \bibnamefont{Huber}},\ }%
  \bibfield{journal}{%
  \Doi{10.1103/PhysRevB.26.3758}{\bibinfo {journal} {Phys. Rev. B}}\ }%
  \textbf{\bibinfo {volume} {26}},\ \bibinfo {pages} {3758} (\bibinfo {year}
  {1982})%
  \bibAnnoteFile{NoStop}{Huber1982}%
\bibitem{GHKDB06}%
  \BibitemOpen
  \bibfield{author}{%
  \bibinfo {author} {\bibfnamefont{K.~Y.}\ \bibnamefont{Guslienko}}, \bibinfo
  {author} {\bibfnamefont{X.~F.}\ \bibnamefont{Han}}, \bibinfo {author}
  {\bibfnamefont{D.~J.}\ \bibnamefont{Keavney}}, \bibinfo {author}
  {\bibfnamefont{R.}~\bibnamefont{Divan}},\ and\ \bibinfo {author}
  {\bibfnamefont{S.~D.}\ \bibnamefont{Bader}},\ }%
  \bibfield{journal}{%
  \Doi{10.1103/PhysRevLett.96.067205}{\bibinfo {journal} {Phys. Rev. Lett.}}\
  }%
  \textbf{\bibinfo {volume} {96}},\ \bibinfo {pages} {067205} (\bibinfo {year}
  {2006})%
  \bibAnnoteFile{NoStop}{GHKDB06}%
\bibitem{UK02}%
  \BibitemOpen
  \bibfield{author}{%
  \bibinfo {author} {\bibfnamefont{N.}~\bibnamefont{Usov}}\ and\ \bibinfo
  {author} {\bibfnamefont{L.}~\bibnamefont{Kurkina}},\ }%
  \bibfield{journal}{%
  \Doi{10.1016/S0304-8853(01)01363-4}{\bibinfo {journal} {{J. Magn. Magn.
  Mater.}}}\ }%
  \textbf{\bibinfo {volume} {242--245 (2)}},\ \bibinfo {pages} {1005} (\bibinfo
  {year} {2002})%
  \bibAnnoteFile{NoStop}{UK02}%
\bibitem{TCCGBT08}%
  \BibitemOpen
  \bibfield{author}{%
  \bibinfo {author} {\bibfnamefont{O.~A.}\ \bibnamefont{Tretiakov}}, \bibinfo
  {author} {\bibfnamefont{D.}~\bibnamefont{Clarke}}, \bibinfo {author}
  {\bibfnamefont{G.-W.}\ \bibnamefont{Chern}}, \bibinfo {author}
  {\bibfnamefont{Y.~B.}\ \bibnamefont{Bazaliy}},\ and\ \bibinfo {author}
  {\bibfnamefont{O.}~\bibnamefont{Tchernyshyov}},\ }%
  \bibfield{journal}{%
  \Doi{10.1103/PhysRevLett.100.127204}{\bibinfo {journal} {Phys. Rev. Lett.}}\
  }%
  \textbf{\bibinfo {volume} {100}},\ \bibinfo {pages} {127204} (\bibinfo {year}
  {2008})%
  \bibAnnoteFile{NoStop}{TCCGBT08}%
\bibitem{ISMW98}%
  \BibitemOpen
  \bibfield{author}{%
  \bibinfo {author} {\bibfnamefont{B.~A.}\ \bibnamefont{Ivanov}}, \bibinfo
  {author} {\bibfnamefont{H.~J.}\ \bibnamefont{Schnitzer}}, \bibinfo {author}
  {\bibfnamefont{F.~G.}\ \bibnamefont{Mertens}},\ and\ \bibinfo {author}
  {\bibfnamefont{G.~M.}\ \bibnamefont{Wysin}},\ }%
  \bibfield{journal}{%
  \Doi{10.1103/PhysRevB.58.8464}{\bibinfo {journal} {Phys. Rev. B}}\ }%
  \textbf{\bibinfo {volume} {58}},\ \bibinfo {pages} {8464} (\bibinfo {year}
  {1998})%
  \bibAnnoteFile{NoStop}{ISMW98}%
\bibitem{ZIPC05}%
  \BibitemOpen
  \bibfield{author}{%
  \bibinfo {author} {\bibfnamefont{C.~E.}\ \bibnamefont{Zaspel}}, \bibinfo
  {author} {\bibfnamefont{B.~A.}\ \bibnamefont{Ivanov}}, \bibinfo {author}
  {\bibfnamefont{J.~P.}\ \bibnamefont{Park}},\ and\ \bibinfo {author}
  {\bibfnamefont{P.~A.}\ \bibnamefont{Crowell}},\ }%
  \bibfield{journal}{%
  \Doi{10.1103/PhysRevB.72.024427}{\bibinfo {journal} {Phys. Rev. B}}\ }%
  \textbf{\bibinfo {volume} {72}},\ \bibinfo {pages} {024427} (\bibinfo {year}
  {2005})%
  \bibAnnoteFile{NoStop}{ZIPC05}%
\bibitem{Doering48}%
  \BibitemOpen
  \bibfield{author}{%
  \bibinfo {author} {\bibfnamefont{W.}~\bibnamefont{D{\"o}ring}},\ }%
  \bibfield{journal}{%
  \bibinfo {journal} {Z. Naturforschung}\ }%
  \textbf{\bibinfo {volume} {3a}},\ \bibinfo {pages} {373} (\bibinfo {year}
  {1948})%
  \bibAnnoteFile{NoStop}{Doering48}%
\bibitem{Ritz09}%
  \BibitemOpen
  \bibfield{author}{%
  \bibinfo {author} {\bibfnamefont{W.}~\bibnamefont{Ritz}},\ }%
  \bibfield{journal}{%
  \bibinfo {journal} {J. Mathematik}\ }%
  \textbf{\bibinfo {volume} {135}},\ \bibinfo {pages} {1} (\bibinfo {year}
  {1909})%
  \bibAnnoteFile{NoStop}{Ritz09}%
\bibitem{Skrotskii84}%
  \BibitemOpen
  \bibfield{author}{%
  \bibinfo {author} {\bibfnamefont{G.~V.}\ \bibnamefont{Skrotskii}},\ }%
  \bibfield{journal}{%
  \bibinfo {journal} {Sov. Phys. Usp.}\ }%
  \textbf{\bibinfo {volume} {27}},\ \bibinfo {pages} {977} (\bibinfo {year}
  {1984})%
  \bibAnnoteFile{NoStop}{Skrotskii84}%
\bibitem{Schlomann73}%
  \BibitemOpen
  \bibfield{author}{%
  \bibinfo {author} {\bibfnamefont{E.}~\bibnamefont{Schl{\"o}mann}},\ }%
  \bibfield{journal}{%
  \Doi{10.1063/1.1653915}{\bibinfo {journal} {Appl. Phys. Lett.}}\ }%
  \textbf{\bibinfo {volume} {19}},\ \bibinfo {pages} {274} (\bibinfo {year}
  {1971})%
  \bibAnnoteFile{NoStop}{Schlomann73}%
\bibitem{Choe2004}%
  \BibitemOpen
  \bibfield{author}{%
  \bibinfo {author} {\bibfnamefont{S.-B.}\ \bibnamefont{Choe}}, \bibinfo
  {author} {\bibfnamefont{Y.}~\bibnamefont{Acremann}}, \bibinfo {author}
  {\bibfnamefont{A.}~\bibnamefont{Scholl}}, \bibinfo {author}
  {\bibfnamefont{A.}~\bibnamefont{Bauer}}, \bibinfo {author}
  {\bibfnamefont{A.}~\bibnamefont{Doran}}, \bibinfo {author}
  {\bibfnamefont{J.}~\bibnamefont{St{\"o}hr}},\ and\ \bibinfo {author}
  {\bibfnamefont{H.~A.}\ \bibnamefont{Padmore}},\ }%
  \bibfield{journal}{%
  \Doi{10.1126/science.1095068}{\bibinfo {journal} {Science}}\ }%
  \textbf{\bibinfo {volume} {304}},\ \bibinfo {pages} {420} (\bibinfo {year}
  {2004})%
  \bibAnnoteFile{NoStop}{Choe2004}%
\bibitem{MG02_JEMS}%
  \BibitemOpen
  \bibfield{author}{%
  \bibinfo {author} {\bibfnamefont{K.~L.}\ \bibnamefont{Metlov}}\ and\ \bibinfo
  {author} {\bibfnamefont{K.~Y.}\ \bibnamefont{Guslienko}},\ }%
  \bibfield{journal}{%
  \bibinfo {journal} {{J. Magn. Magn. Mater.}}\ }%
  \textbf{\bibinfo {volume} {242--245}},\ \bibinfo {pages} {1015} (\bibinfo
  {year} {2002})%
  \bibAnnoteFile{NoStop}{MG02_JEMS}%
\bibitem{Gilbert55}%
  \BibitemOpen
  \bibfield{author}{%
  \bibinfo {author} {\bibfnamefont{T.}~\bibnamefont{Gilbert}},\ }%
  \bibfield{journal}{%
  \bibinfo {journal} {Phys. Rev.}\ }%
  \textbf{\bibinfo {volume} {100}},\ \bibinfo {pages} {1243} (\bibinfo {year}
  {1955})%
  \bibAnnoteFile{NoStop}{Gilbert55}%
\bibitem{Kambersky2007}%
  \BibitemOpen
  \bibfield{author}{%
  \bibinfo {author} {\bibfnamefont{V.}~\bibnamefont{Kambersk{\'y}}},\ }%
  \bibfield{journal}{%
  \Doi{10.1103/PhysRevB.76.134416}{\bibinfo {journal} {Phys. Rev. B}}\ }%
  \textbf{\bibinfo {volume} {76}},\ \bibinfo {pages} {134416} (\bibinfo {year}
  {2007})%
  \bibAnnoteFile{NoStop}{Kambersky2007}%
\bibitem{suppfrequency}%
  \BibitemOpen
  \bibinfo {note} {See Supplemental Material for Mathematica code computing
  potential energy expansion coefficients and vortex precession frequency.}%
  \bibAnnoteFile{Stop}{suppfrequency}%
\bibitem{MG04}%
  \BibitemOpen
  \bibfield{author}{%
  \bibinfo {author} {\bibfnamefont{K.~L.}\ \bibnamefont{Metlov}}\ and\ \bibinfo
  {author} {\bibfnamefont{K.~Y.}\ \bibnamefont{Guslienko}},\ }%
  \bibfield{journal}{%
  \bibinfo {journal} {{Phys. Rev. B}}\ }%
  \textbf{\bibinfo {volume} {70}},\ \bibinfo {pages} {052406} (\bibinfo {year}
  {2004})%
  \bibAnnoteFile{NoStop}{MG04}%
\bibitem{M06}%
  \BibitemOpen
  \bibfield{author}{%
  \bibinfo {author} {\bibfnamefont{K.~L.}\ \bibnamefont{Metlov}},\ }%
  \bibfield{journal}{%
  \bibinfo {journal} {{Phys. Rev. Lett.}}\ }%
  \textbf{\bibinfo {volume} {97}},\ \bibinfo {pages} {127205} (\bibinfo {year}
  {2006})%
  \bibAnnoteFile{NoStop}{M06}%
\bibitem{Feynman1949}%
  \BibitemOpen
  \bibfield{author}{%
  \bibinfo {author} {\bibfnamefont{R.~P.}\ \bibnamefont{Feynman}},\ }%
  \bibfield{journal}{%
  \Doi{10.1103/PhysRev.76.749}{\bibinfo {journal} {Phys. Rev.}}\ }%
  \textbf{\bibinfo {volume} {76}},\ \bibinfo {pages} {749} (\bibinfo {year}
  {1949})%
  \bibAnnoteFile{NoStop}{Feynman1949}%
\end{thebibliography}
%Merlin.mbs v4.21 2009-07-09.
%

\end{document}